\documentclass{emulateapj}
\usepackage{graphicx}

\shorttitle{A clumpy, rotating gas disk in SMG GN20}
\shortauthors{Hodge et al.}

\begin{document}

\title{Evidence for a clumpy, rotating gas disk in a Submillimeter Galaxy at z = 4}

\author{J. A. Hodge\altaffilmark{1}}
\altaffiltext{1}{Max-Planck Institute for Astronomy, K\"{o}nigstuhl 17, 69117 Heidelberg, Germany}
\email{hodge@mpia.de}

\author{C. L. Carilli\altaffilmark{2,3}}
\altaffiltext{2}{National Radio Astronomy Observatory, P.O. Box 0, Socorro, NM 87801-0387, USA}
\altaffiltext{3}{Astrophysics Group, Cavendish Laboratory, JJ Thomson Avenue, Cambridge CB3 0HE, UK}

\author{F. Walter\altaffilmark{1,2}}

\author{W. J. G. de Blok\altaffilmark{4,5}}
\altaffiltext{4}{ASTRON, PO Box 2, 7990 AA Dwingeloo, The Netherlands}
\altaffiltext{5}{Astrophysics, Cosmology and Gravity Centre (ACGC), Astronomy Department, University of Cape Town, Private Bag X3, 7700 Rondebosch, Republic of South Africa}

\author{D. Riechers\altaffilmark{6}}
\altaffiltext{6}{Department of Astronomy, California Institute of Technology, MC 249-17, 1200 East California Boulevard, Pasadena, CA 91125, USA}

\author{E. Daddi\altaffilmark{7}}
\altaffiltext{7}{CEA, Laboratoire AIM-CNRS-Universit\'{e} Paris Diderot, Irfu/SAp, Orme des Merisiers, F-91191 Gif-sur-Yvette, France}

\author{L. Lentati\altaffilmark{3}}

\begin{abstract}
We present Karl G. Jansky Very Large Array (VLA) observations of the CO(2--1) emission in the $z=4.05$ submillimeter galaxy (SMG) GN20. These high--resolution data allow us to image the molecular gas at 1.3 kpc resolution just 1.6 Gyr after the Big Bang. The data reveal a clumpy, extended gas reservoir, 14 $\pm$ 4 kpc in diameter, in unprecedented detail. A dynamical analysis shows that the data are consistent with a rotating disk of total dynamical mass 5.4 $\pm$ 2.4 $\times$ 10$^{11}$M$_{\odot}$. We use this dynamical mass estimate to constrain the CO--to--H$_2$ mass conversion factor ($\alpha_{\rm CO}$), finding $\alpha_{\rm CO}=1.1\pm0.6$ M$_{\sun}$(K km s$^{-1}$ pc$^{2})^{-1}$. We identify five distinct molecular gas clumps in the disk of GN20 with masses a few percent of the total gas mass, brightness temperatures of 16--31K, and surface densities of $>3,200-4,500\times(\alpha_{\rm CO}/0.8$) M$_{\odot}$ pc$^{-2}$.  Virial mass estimates indicate they could be self--gravitating, and we constrain their CO--to--H$_2$ mass conversion factor to be $<$0.2--0.7 M$_{\sun}$(K km s$^{-1}$ pc$^{2})^{-1}$. A multiwavelength comparison demonstrates that the molecular gas is concentrated in a region of the galaxy that is heavily obscured in the rest--frame UV/optical. We investigate the spatially--resolved gas excitation and find that the CO(6--5)/CO(2--1) ratio is constant with radius, consistent with star formation occuring over a large portion of the disk. We discuss the implications of our results in the context of different fueling scenarios for SMGs.


\noindent\textit{Key words:} galaxies: evolution $--$ galaxies: formation $--$ galaxies: high-redshift $--$ galaxies: ISM $--$ galaxies: star formation

\end{abstract}
\section{INTRODUCTION}
\label{Intro}

Submillimeter--luminous galaxies \citep[SMGs;][]{2002PhR...369..111B} are dusty, gas--rich, high--z galaxies that were revealed in the first extragalactic surveys using SCUBA and MAMBO \citep[e.g.,][]{1997ApJ...490L...5S, 1998Natur.394..241H, 1998Natur.394..248B, 1999ApJ...515..518E, 1999MNRAS.302..632B, 2000A&A...360...92B, 2004MNRAS.354..779G}.
Their huge far--infrared (FIR) luminosities ($\sim$10$^{13}$ $L_{\odot}$) are believed to be primarily driven by intense ($\sim$10$^{3}M_{\odot}$ yr$^{-1}$) star formation \citep[e.g.,][]{2005ApJ...632..736A, 2003AJ....125..383A}, adding to their already significant stellar masses \citep[e.g.,][]{2005ApJ...635..853B,2011ApJ...740...96H}.  
Together with FIR luminous quasars, they are generally thought to evolve into the giant ellipticals we see in the local universe.  They are therefore critical for understanding early--type massive galaxy formation.


SMGs are a relatively rare phenomenon, with typical space densities of 10$^{-5}$--10$^{-6}$ Mpc$^{-3}$. 
Their rarity may be partly because their 
enhanced SFRs are necessarily short--lived \citep[$<$ 100 Myr;][]{2005MNRAS.359.1165G}, 
or else the galaxies would grow too large, too fast.
The highest volume density of radio--selected SMGs occurs at $z \sim 2-3$, 
indicating that they peak simultaneously with the peak epoch of star formation \citep{2003Natur.422..695C}. 
However, a number of recently discovered higher--redshift SMGs \citep{2008ApJ...689L...5S, 2009ApJ...694.1517D, 2009ApJ...695L.176D, 2010ApJ...720L.131R, 2010MNRAS.407L.103C, 2011Natur.470..233C, 2011MNRAS.415.1479W, 2011ApJ...740...63C} 
as well as several studies based on statistical arguments \citep[][and references therein]{2008MNRAS.389.1489G, 2011MNRAS.410.2749P}
may indicate that a high--redshift tail does exist
and is able to account for the population of old, massive ellipticals already in place at $z \sim 2-3$ \citep{2009ApJ...694.1517D}.

Many SMGs are believed to be starburst--dominated major mergers \citep[e.g.,][]{2003ApJ...599...92C, 2010ApJ...724..233E,2010MNRAS.401.1613N,2011ApJ...743..159H,2012arXiv1203.1318H}.
This would make SMGs the high--redshift analogs of Ultra--Luminous Infrared Galaxies (ULIRGS) in the local universe.
Indeed, there is direct evidence for multiple CO components and/or disturbed kinematics in some SMGs, supporting the merger picture \citep[e.g.,][]{2008ApJ...680..246T, 2010ApJ...724..233E, 2011MNRAS.412.1913I, 2011ApJ...739L..31R}. 


Recently, it has been suggested that other mechanisms that drive extreme star formation rates may also be at play. 
In particular, a scenario has been put forward in which the star formation in massive, high--redshift galaxies is driven by cold mode accretion \citep[CMA; e.g.][]{2005MNRAS.363....2K, 2009Natur.457..451D, 2009ApJ...703..785D}.  
CMA--driven galaxies are constantly forming stars at high rates, and the star formation is sustained by smooth infall and accretion of gas--rich material.  
This process can result in elliptical galaxies because the streams that feed star formation can cause the disk to break up into giant clumps if they have a high enough gas fraction and degree of turbulence. 
The clumps then potentially migrate inward and merge into a spheroid \citep{2009Natur.457..451D, 2009ApJ...703..785D}. 

The CMA phenomenon has been extended to SMGs by a number of authors \citep{2001astro.ph..7290F, 2006ApJ...639..672F, 2010MNRAS.404.1355D,2010ApJ...714.1407C}. 
In this scenario, SMGs are massive galaxies sitting at the centers of deep potential wells and fed by smooth accretion.  
They can be thought of as super--sized versions of normal star forming galaxies, providing an alternate formation scenario to the merger-induced model.
By identifying simulated SMGs as the most rapidly star--forming systems that match the number densities of SMGs, this theory has been successful at explaining some key SMG properties, including their stellar masses and clustering scales \citep{2010MNRAS.404.1355D}.


According to \citet{2008ApJ...682..231S}, the best way to distinguish between CMA and a gas--rich merger model is through observations of the gas dynamics and distribution.  
A well defined disk with a smooth rotation curve would be indicative of CMA, while tidally disturbed gas, with a very high T$_{\rm B}$ starburst nucleus, would point towards a merger.  
This is simply due to the redistribution of angular momentum.
The large--scale gravitational torques induced by gas--rich major mergers
are efficient at removing angular momentum \citep{1996ApJ...471..115B},
thereby funneling the cold molecular gas into the galaxy's center and producing a nuclear starburst.
Indeed, in ULIRGs -- the canonical merger scenario -- the gas is concentrated in the central kpc \citep[e.g.,][]{1998ApJ...507..615D, 1999AJ....117.2632B}.
The gas--rich, star--forming disks in Arp220 have a size of only $\sim$100 pc  \citep{2008ApJ...684..957S}.
In simulations where SMGs result from mergers, their gas reservoirs show slightly larger extents of a few kpc \citep{2009MNRAS.400.1919N}.

Conversely, the $z\sim2$ massive star forming galaxies (SFGs) which have been proposed to be due to smooth accretion are large disk galaxies, with star formation occuring at large radii or situated in rings \citep{2006ApJ...651..676E, 2008ApJ...687...59G, 2009ApJ...692...12E}. 
Rest--frame UV/optical imaging shows that, unlike low--redshift galaxies, their disks tend to be broken into multiple giant clumps of $\sim$1 kpc and 10$^{9}$ M$_{\odot}$ \citep{2004ApJ...604L..21E, 2005ApJ...627..632E, 2006ApJ...645.1062F, 2008ApJ...687...59G}. 
More recent work has also imaged such galaxies in millimeter/CO emission
\citep{2010Natur.463..781T, 2010Natur.464..733S, 2011ApJ...742...11S},
finding evidence for clumpy CO emission extending over the disk \citep[e.g.][]{2010Natur.463..781T}.

Observations of the morphology and kinematics of molecular gas in SMGs may therefore shed light on the physics behind the intense star formation.
Previously, \citet{2010ApJ...714.1407C} presented VLA observations of CO(1--0) and CO(2--1) emission in the $z = 4.05$ SMG GN20, 
the brightest SMG in the GOODS--N field \citep{2006MNRAS.370.1185P}
and one of three SMGs in what appears to be a massive $z\sim4$ proto--cluster \citep{2009ApJ...694.1517D}.
These low--J transitions are of particular interest because they trace the cold molecular gas thought to make up the bulk of the molecular gas in these systems.
Although their high--resolution CO(2--1) imaging showed evidence that the gas in GN20 was well--resolved, their observations suffered from some severe spectral limitations.
In particular, the observations utilized the old VLA correlator, with a total bandwidth of only 100 MHz (650 km s$^{-1}$), causing the line profiles to be truncated on both sides.
The observations were also done in continuum mode, resulting in no information on kinematics.

We therefore obtained over 120 hours of time on the Karl G. Jansky Very Large Array \citep[VLA;][]{2011arXiv1106.0532P} to image the CO(2--1) emission in the GN20 field in the B-- and D--configurations. 
The $\sim$20 hours of lower--resolution D--array data were presented by \citet{2011ApJ...739L..33C}. 
Here, we present the full dataset (B+D--configurations) on GN20, providing greatly improved spatial resolution and image fidelity. 

We begin in Section~\ref{obs} by describing our new VLA observations and data reduction of GN20. CO maps and the derived gas mass are presented in Section~\ref{results}. 
Section~\ref{analysis} constitutes our analysis, 
including 
a dynamical analysis (\ref{dynanal}), 
the definition and properties of individual gas clumps (\ref{clumps}), 
a multiwavelength comparison (\ref{multiwav}), 
and an analysis of the spatially--resolved gas excitation (\ref{excitation}).
We discuss the implications of our results on the nature of GN20 in
Section~\ref{discussion}, and 
we end with a summary of our conclusions in Section~\ref{conclusions}.  
Where applicable we assume the standard $\Lambda$ cosmology of H$_0$ = 70 km s$^{-1}$ Mpc$^{-1}$, $\Omega_{\Lambda}$ = 0.7, and $\Omega_{M}$ = 0.3 \citep{2003ApJS..148..175S, 2007ApJS..170..377S}.
At a redshift of $z = 4.05$, 1$^{\prime\prime}$ corresponds to 7 kpc.

\begin{figure}[]
\centering
\includegraphics[scale=0.5]{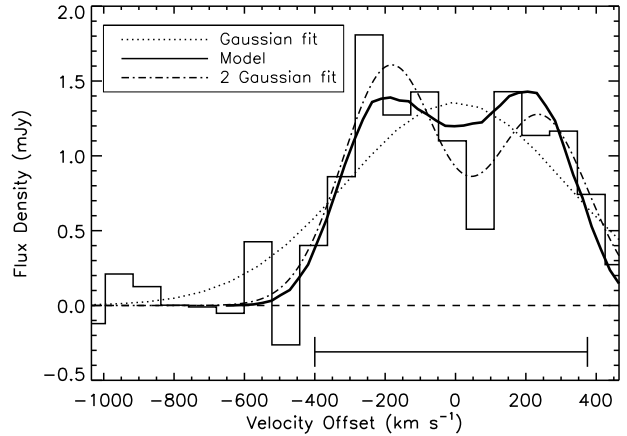}
\caption{CO(2--1) spectrum of GN20, binned into channels of 78 km s$^{-1}$.  The single Gaussian fit to the data is shown by the dotted line, and the double--horn fit is shown by the dot--dashed line. The bar at the bottom of the plot shows the velocity range averaged over to make Figure~\ref{fig:GN20_120MHz}. The model fit results from the dynamical modeling discussed in Section~\ref{dyn}.}
\label{fig:GN20spec}
\end{figure}

\begin{figure*}[]
\centering
\includegraphics[scale=0.65]{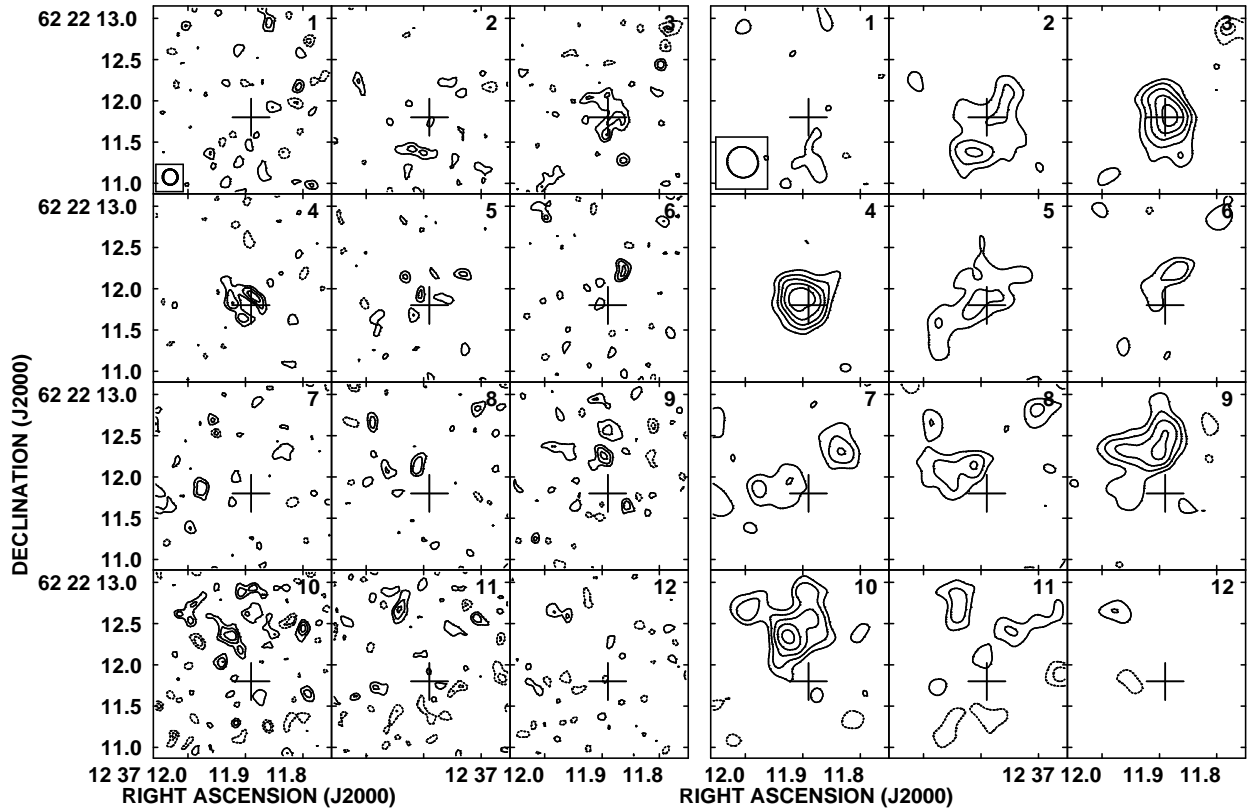}
\caption{VLA CO(2--1) in GN20 in 78 km s$^{-1}$ channels. Increasing channel numbers indicate increasing frequency (decreasing velocity), and channel 6 corresponds to the central frequency derived from the spectrum (Figure~\ref{fig:GN20spec}). The left panel is at 0.19$^{\prime\prime}$ resolution, and the right panel shows the same channels tapered to 0.38$^{\prime\prime}$ resolution to increase the S/N ratio. The cross shows the peak of the 1.4 GHz counterpart at 1.7$^{\prime\prime}$ resolution \citep{2010ApJS..188..178M}.  The rms noise values are 59 $\mu$Jy beam$^{-1}$ and 70 $\mu$Jy beam$^{-1}$, and the contours are given in steps of 1$\sigma$ starting at $\pm$2$\sigma$.}
\label{fig:GN20_ChMaps}
\end{figure*}

\section{OBSERVATIONS \& DATA REDUCTION}
\label{obs}

We observed the CO(2--1) transition toward the GN20 field as part of VLA key project AC974. 
The project was awarded 96 hours in B--configuration (baselines up to 10 km) and 28 hours in D--configuration (baselines up to 1 km), for a total of 124 hours.
The observations were dynamically scheduled and took place in March--April 2010 (D--configuration), and February--April 2011 (B--configuration).  

The CO(2--1) line (rest frequency $\nu$ $=$ 230.5424 GHz) is redshifted to $\nu$ $=$ 45.655 GHz at $z = 4.05$, requiring the Q--band.  
The primary beam is $\sim$1$^{\prime}$ (FWHM) at this frequency, and the pointing center was chosen to be 10$^{\prime\prime}$ west of GN20 
so that GN20, the nearby SMGs (and fellow proto--cluster members) GN20.2a and GN20.2b, and the $z=1.5$ galaxy BzK--21000 would all fall within the 70\% sensitivity radius of the primary beam (data presented in \citealt{Hodge_gn20envi}).
All images have been corrected for the response of the VLA primary beam. 
We centered the two 128 MHz IFs at 45.592 GHz and 45.720 GHz, for a total bandwidth of 246 MHz or 1600 km s$^{-1}$ (taking into account the overlap).  
Each of the two IFs had 64 channels, 
resulting in an instrumental velocity resolution of 13 km s$^{-1}$.  
Observations were taken in full polarization mode.

We used fast switching phase calibration \citep{1999RaSc...34..817C}, with a cycle time of 5 minutes.  VLA calibrator J1302+5748 served as the phase calibrator.  The quasar 3C286 was used to determine the absolute flux density scale and for bandpass calibration.  The B--configuration data were reduced using the Astronomical Image Processing System (AIPS), and the D--configuration data were reduced in the Common Astronomy Software Applications (CASA) package.  During the reduction, we discarded data taken during times of poor phase stability. After accounting for calibration overheads and flagging, the total time on source was approximately 50 hours. The data were then combined within AIPS for further analysis.  

We imaged the data using the CLEAN algorithm, and we cleaned down to 1.5$\sigma$ in a 
2$^{\prime\prime}$ $\times$ 2$^{\prime\prime}$ clean box around GN20. 
A spectrum was extracted using an aperture the same size as the clean box.  
The best compromise between resolution and sensitivity for this dataset resulted from using Briggs weighting with a robust parameter of R $=$ 1.0.
This results in an angular resolution of 0.19$^{\prime\prime}$, or 1.3 kpc at the redshift of GN20.
We will refer to this resolution as our `native' resolution for the remainder of the paper.



\section{RESULTS}
\label{results}

\subsection{CO(2--1) Detection and Channel Maps}
The CO(2--1) spectrum for GN20 is shown in Figure~\ref{fig:GN20spec}.
We used a Gaussian fit to the spectrum at 12 MHz (78 km s$^{-1}$) resolution to derive a peak flux density of 1.35 $\pm$ 0.20 mJy and a FWHM of 730 $\pm$ 140 km s$^{-1}$, 
consistent with \citet{2011ApJ...739L..33C}. 
The integrated flux is 1.0 $\pm$ 0.25 Jy km s$^{-1}$ (see Table~\ref{tab-1}).
The spectrum indicates a redshift for the galaxy of z $=$ 4.0548 (topocentric), consistent with earlier measurements, including high--J transitions \citep{2009ApJ...694.1517D}.
We detect no continuum emission from GN20 at our native resolution (0.19$^{\prime\prime}$ with a 3$\sigma$ limit of 39.5 $\mu$Jy
and at 1.75$^{\prime\prime}$ resolution with a 3$\sigma$ limit of 65.4 $\mu$Jy.

The spectrum shows the clear indication of a double--peaked, or at least flattened, structure, the signature of which is also evident in a position--velocity diagram created at a later point in the analysis (Section~\ref{dyn}). 
If we instead fit the spectrum with a combination of two Gaussians (Figure~\ref{fig:GN20spec}; dot--dashed line), we find that the two components have peak flux densities of 1.6 $\pm$ 0.3 mJy and 1.3 $\pm$ 0.3 mJy, and FWHM values of 300 $\pm$ 90 km s$^{-1}$ and 320 $\pm$ 130 km s$^{-1}$, respectively. 
The integrated flux of the combined components is 1.0 $\pm$ 0.3 Jy km s$^{-1}$, consistent with the result from the single--Gaussian fit above.
We will use the two--Gaussian fit 
to determine the CO luminosity and gas mass	
in the remaining analysis.

Channel maps of 78 km s$^{-1}$ width are shown in Figure~\ref{fig:GN20_ChMaps}. The left panel shows the emission at our native resolution (0.19$^{\prime\prime}$), and the right panel shows the same channels tapered to 0.38$^{\prime\prime}$ resolution in order to increase the S/N ratio.
The two outer channels in each case sample the continuum on either side of the line.  The emission first appears 
in the South of the plotted field,
 and it shifts to the North with increasing frequency.  
We will analyze the kinematics of the system in Section~\ref{dynanal} below. 

\begin{deluxetable}{ l c }
\tabletypesize{\small}
\tablewidth{0pt}
\tablecaption{GN20 Observed and Derived Parameters \label{tab-1}}
\tablehead{
\colhead{\textbf{Parameter}} & \colhead{\textbf{Value}}}
\startdata
Position (J2000)\tablenotemark{a} & 12$^{h}$37$^{m}$11.89$^{s}$ +62$^{\circ}$22$^{\prime}$11.8$^{\prime\prime}$ \\
z & 4.0548 $\pm$ 0.0008 \\
S$_{\rm CO(2-1)}$ & 1.6 $\pm$ 0.3 mJy (peak 1) \\
 & 1.3 $\pm$ 0.3 mJy (peak 2) \\
FWHM$_{\rm CO(2-1)}$ & 300 $\pm$ 90 km s$^{-1}$ (peak 1) \\
 & 320 $\pm$ 130 km s$^{-1}$ (peak 2) \\
$I_{\rm CO(2-1)}$ & 1.0 $\pm$ 0.3 Jy km s$^{-1}$  \\
$L^{\prime}_{\rm CO(2-1)}$  &  1.6 $\pm$ 0.5 $\times$ 10$^{11}$ K km s$^{-1}$ pc$^{2}$ \\
$M(\rm H_2)$ & 1.3 $\pm$ 0.4 $\times$ 10$^{11}$ $\times$ $(\alpha_{\rm CO}/0.8)$ M$_{\sun}$ \\
$M_{\rm dyn}$ &  5.4 $\pm$ 2.4 $\times$ 10$^{11}$ M$_{\sun}$ 
\enddata
\tablenotetext{a}{From the 1.4 GHz observations of \citet{2010ApJS..188..178M} at 1.7$^{\prime\prime}$ resolution. All other parameters are from the study presented here.}
\end{deluxetable}

\subsection{Molecular Gas Mass}
\label{H2mass}

From the two--Gaussian fit to the spectrum of GN20, we derive a CO luminosity of $L^{\prime}_{\rm CO(2-1)}$ $=$ 1.6 $\pm$ 0.5 $\times$ 10$^{11}$ K km s$^{-1}$ pc$^{2}$.
This implies a molecular gas mass of $M(\rm H_2)$ $=$ 1.3 $\pm$ 0.4 $\times$ 10$^{11}$ $\times$ $(\alpha_{\rm CO}/0.8)$ M$_{\sun}$ (see Table 1) 
 assuming the standard relationships from \citet{2005ARA&A..43..677S}.  
As justified by \citet{2010ApJ...714.1407C}, we assumed thermal excitation for the extrapolation from CO(2--1) to CO(1--0), and we used a CO--to--H$_2$ conversion factor of $\alpha_{\rm CO}$ $=$ 
0.8 M$_{\sun}$ (K km s$^{-1}$ pc$^{2}$)$^{-1}$, 
the value typically assumed for ULIRGs and SMGs \citep{1998ApJ...507..615D, 2005ARA&A..43..677S, 2006ApJ...640..228T, 2008ApJ...680..246T}.  
In Section~\ref{Mdyn} below, we use our dynamical mass estimate to put constraints on $\alpha_{\rm CO}$ for this system.

\begin{figure}[]
\centering
\includegraphics[scale=0.75]{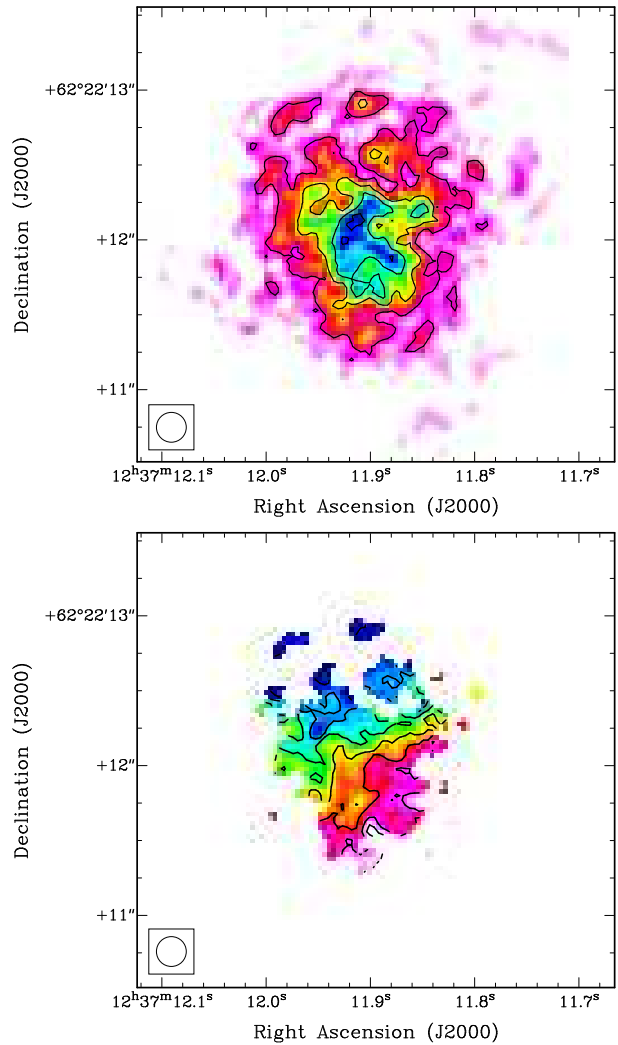}
\caption{CO(2--1) 0th (top) and 1st (bottom) moment maps for GN20 at a resolution of 0.19$^{\prime\prime}$. 
The 0th moment map (i.e. integrated intensity) has a peak S/N of ~6, and the contours shown start at (and are in steps of) 15.5 mJy km s$^{-1}$. 
Contours for the 1st moment map (i.e., intensity--weighted mean velocity) are shown for steps of 100 km s$^{-1}$, with the contour between the green and orange bands representing the systemic velocity.}
\label{fig:GN20_mom}
\end{figure}


\section{ANALYSIS}
\label{analysis}

\subsection{Dynamical Analysis}
\label{dynanal}

\subsubsection{Moment Maps}
\label{moms}

Given the high--quality of the data, we created moment maps by following the typical approach used for HI and CO observations of nearby galaxies \citep[e.g.][]{2008AJ....136.2563W, 2009AJ....137.4670L}. 
This approach is useful for high resolution datasets such as this (0.19$^{\prime\prime}$), where real emission can be resolved out in a moment map with a single, global S/N cut. 
To address this fact, this technique involves using the data, tapered to a lower resolution (hence recovering more extended emission), as an additional input. 
For this purpose, we used data tapered to 0.38$^{\prime\prime}$ resolution.
This lower--resolution datacube is used to create a mask (on a per--channel basis) of the significant emission, including more diffuse emission. 
These masks are then used to blank the higher--resolution data, thereby retaining real, diffuse emission in the high--resolution data that would have otherwise been blanked.
This standard procedure is the best possible method to recover diffuse, low S/N emission while retaining the highest possible spatial resolution. 

\begin{figure*}[]
\centering
\includegraphics[scale=0.65]{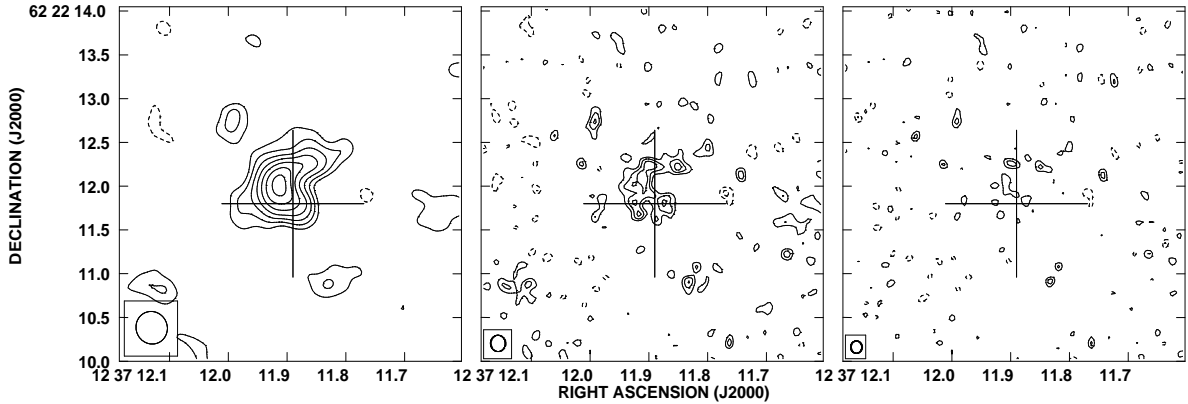}
\caption{Velocity--averaged B$+$D--array CO(2--1) images of GN20 over a bandwidth of 780 km s$^{-1}$ at three different resolutions (from left): 0.38$^{\prime\prime}$, 0.19$^{\prime\prime}$, and 0.14$^{\prime\prime}$, as given in the lower left corner of the maps. The cross shows the peak of the 1.4 GHz counterpart and the size of the cross indicates the 1.4 GHz beam at 1.7$^{\prime\prime}$ resolution \citep{2010ApJS..188..178M}. The rms noise values are 25.6 $\mu$Jy beam$^{-1}$, 19.0 $\mu$Jy beam$^{-1}$, and 25.0 $\mu$Jy beam$^{-1}$, and the contours are given in steps of 1$\sigma$ starting at $\pm$2$\sigma$.}
\label{fig:GN20_120MHz}
\end{figure*}

We used the method described above to create the 0th moment (integrated intensity) map, shown in Figure~\ref{fig:GN20_mom} (top panel).
As the masking process is done on a per--channel basis, the noise in any given pixel in the map is given by $\sqrt{N}\times\sigma_{\rm chan}$, where $N$ is the number of channels that were integrated for that particular pixel, and $\sigma_{\rm chan}$ is the rms noise per channel.
We used this information to apply an additional S/N cut to the 1st moment map (intensity--weighted velocity; Figure~\ref{fig:GN20_mom} lower panel).
In particular, we required S/N $>$ 3 for the 1st moment map, since any remaining unmasked noise will have a large effect on the resulting velocity field.  
Although some noise is still present in the outskirts, a clear velocity gradient is apparent across the disk.

As a verification of the blanking process described above, we show unblanked velocity--averaged maps of GN20 in Figure~\ref{fig:GN20_120MHz} at three different angular resolutions. 
The left panel has been tapered to 0.38$^{\prime\prime}$, the middle panel is at the native resolution (0.19$^{\prime\prime}$), and the right panel uses Briggs weighting with R$=$--0.5 to reach 0.14$^{\prime\prime}$ resolution (1.0 kpc at z $=$ 4.05).
In contrast to the moment analysis described above, these maps were made by simply averaging over 780 km s$^{-1}$ (the region indicated in Figure~\ref{fig:GN20spec}), without first blanking the noise on a per--channel basis.
The gas distribution appears slightly different than that seen in the 0th moment map because the channels without emission are included in the average.  
The highest--resolution map confirms that the emission is not concentrated in just one or two strong peaks, but rather spread out over a larger area.

The total flux density (averaged over 780 km s$^{-1}$) in these velocity--averaged B$+$D--array maps is 633 $\pm$ 67 $\mu$Jy. 
For comparison, the total flux densities in the B--array only (extended configuration) and D--array only (compact configuration) maps (not shown) are 365 $\pm$ 57 $\mu$Jy and 813 $\pm$ 65 $\mu$Jy, respectively. 
The total flux density in the 0th moment map (800 $\pm$ 74 $\mu$Jy) is consistent with that in the D--array only map, confirming that the recovered emission in the 0th moment map is real.
The blanking process described above therefore allows us to achieve the native resolution while still recovering the diffuse emission present on larger scales.

We estimate the size of GN20's gas reservoir from the 0th moment map (Figure~\ref{fig:GN20_mom}, top).
Defining the radius as the maximum radial extent of the resolved CO(2--1) emission in the map, 
and conservatively assuming an uncertainty in the measurement of $\sim$30\%,
we derive a radius of 1$^{\prime\prime}$ $\pm$ 0.3$^{\prime\prime}$, equivalent to $\sim$7 kpc ($\pm$ 2 kpc) at $z = 4.055$.
The total diameter of the source in CO(2--1) is therefore $\sim$14 $\pm$ 4 kpc.
The large extent of the molecular gas reservoir is not unlike what has been seen in some other (lower--z) SMGs in low--excitation imaging;
while SMGs typically have relatively compact distributions in the higher--J CO lines \citep[HWHM$=$2--4 kpc; e.g.][]{2006ApJ...640..228T}, recent observations of SMGs in CO(1--0) show more extended gas reservoirs \citep[e.g.][]{2011MNRAS.412.1913I, 2011ApJ...733L..11R, 2011ApJ...739L..31R, 2010MNRAS.404..198I}.

\begin{figure*}[]
\centering
\includegraphics[scale=0.8]{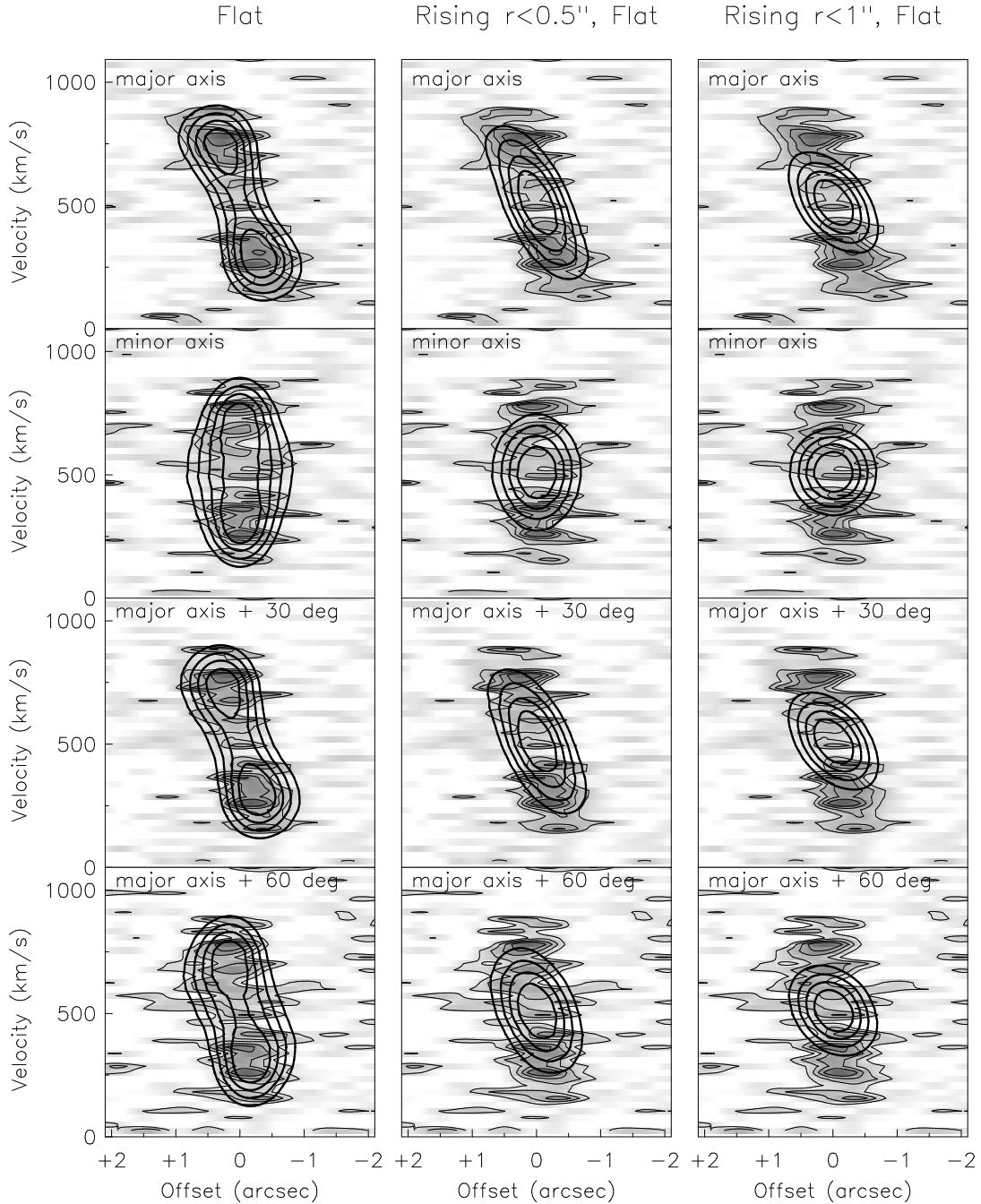}
\caption{Position--velocity diagrams for CO(2--1) emission in GN20 for four different slices and three different models (shown as contours).  The input data cube has a spectral resolution of 0.77$^{\prime\prime}$ and a spectral resolution of 26 km s$^{-1}$. The different slices are shown in the different rows and are (from top): major axis (i.e. a position angle of 25$^{\circ}$), minor axis, major axis$+$30$^{\circ}$, and major axis$+$60$^{\circ}$. The columns show the different models (contours): a flat rotation curve (left), a rotation curve that rises linearly to 0.5$^{\prime\prime}$ (3.5 kpc) then flattens (middle), and a rotation curve that rises linearly out to 1.0$^{\prime\prime}$ (7 kpc) (right). The velocities on the vertical axis are relative.  Greyscale and thin countours show the observed data. The data contours are 55\% to 85\% in steps of 10\%. For all models, the contours are 35\% to 80\% of the peak brightness, in steps of 15\%. The left--hand panel shows our best fit model (see text for details).}
\label{fig:galmod_rv}
\end{figure*}

\subsubsection{Dynamical Modeling}
\label{dyn}
For the dynamical modeling of GN20, we used the GALMOD task (part of the GIPSY package).
GALMOD creates a three-dimensional model using a set of input parameters, and it then convolves the model cube to the spatial/spectral resolution of the data for comparison.
We used an input data cube with a spectral resolution of 26 km s$^{-1}$, and we tapered the data to an angular resolution of 0.77$^{\prime\prime}$ as it was found that higher resolutions resolved out too much emission to be usable. 
The GALMOD task requires a radial profile as input, which (guided by the 0th moment map) we set as an exponential radial profile with a slope of --0.4.
We used a thin disk model
and we found that changes in the thickness of the disk (within a reasonable range, $<$ few kpc) did not result in major changes to the model. 

As the S/N of our data have necessitated using a lower resolution (0.77$^{\prime\prime}$) cube for the modeling, it is not possible to measure the intrinsic rotation curve of this source directly from the data.
Instead, we have assumed a rotation curve as input and then compared the model to the data in position--velocity space. 
Position--velocity diagrams are shown for three different input rotation curves in Figure~\ref{fig:galmod_rv}.
The columns represent the different rotation curves: The left--hand column assumes a flat rotation curve, the middle column assumes a rotation curve that rises linearly from 0$^{\prime\prime}$--0.5$^{\prime\prime}$ (0--3.5 kpc) to v$_{\rm max}$, then is flat at larger radii, and the right--hand column assumes a rotation curve that rises linearly from 0$^{\prime\prime}$--1.0$^{\prime\prime}$ to v$_{\rm max}$ (0--7 kpc), then is flat at larger radii.
The rows show four different slices through the three--dimensional data: major axis, taken at a position angle of 25$^{\circ}$ (top row), minor axis (second row), major axis $+$ 30$^{\circ}$ (third row), and major axis $+$ 60$^{\circ}$ (bottom row). 
These comparison plots demonstrate that, while the resolution of the input data cube (due to our limited S/N) make it difficult to constrain the exact shape of the rotation curve, we can constrain the curve to having reached its flat part at least within 0.5$^{\prime\prime}$ (3.5 kpc), with a preference for the flat rotation curve.

\begin{figure*}[]
\centering
\includegraphics[scale=0.7]{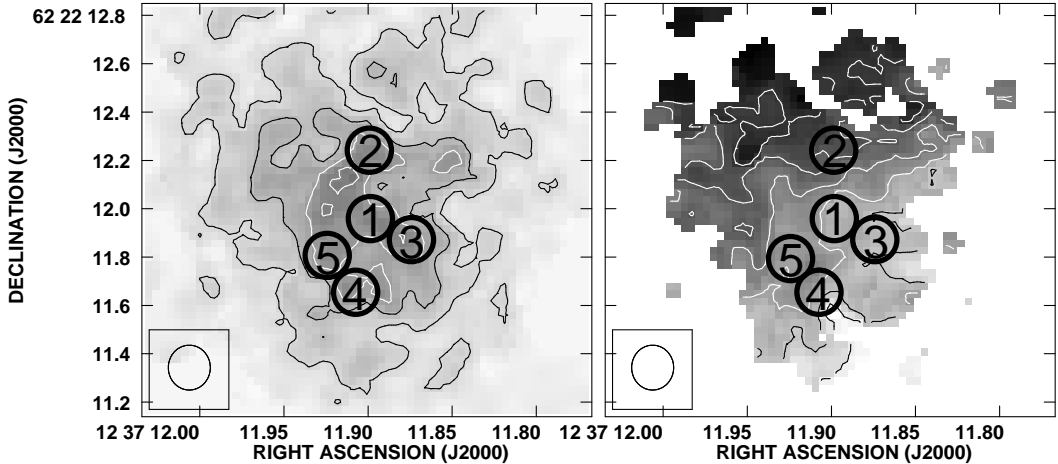}
\caption{Zoomed--in CO(2--1) 0th (left) and 1st (right) moment maps at 0.19$^{\prime\prime}$ resolution. The clumps in the final clump sample are indicated by the numbered circles. }
\label{fig:GN20_clumps}
\end{figure*}

We therefore find that the velocity field is fully consistent with a rotating disk
with a flat rotation curve (as discussed above).
By comparing different models to the data, we find that the best--fit model is a rotating disk with an inclination of $i =$ 30$^{\circ}$ $\pm$ 15$^{\circ}$, 
a maximum rotational velocity of v$_{\rm max}$ $=$ 575 $\pm$ 100 km s$^{-1}$, and a dispersion of $\delta = 100$ $\pm$ 30 km s$^{-1}$.  
Note that the error bars are not statistical, but are liberal estimates of the uncertainty determined through careful comparison between model and data.
Note also that deriving the dispersion from a spatially and spectrally convolved disk model, unlike other mean--weighted dispersion estimators, is unbiased by beam smearing \citep{2011ApJ...741...69D}.
The spectrum predicted by our best--fit model is a good fit to our observed spectrum (Figure~\ref{fig:GN20spec}).

The relatively large value of v$_{\rm max}$ is due to the fairly small inclination value;  while the quantity v$_{\rm max}$sin$(i)$ is well constrained, the two components are difficult to disentangle at our angular resolution. 
Following standard procedure for the modeling of low to medium resolution HI observations, the final inclination value of 30$^{\circ}$ was chosen to reproduce the resulting ellipticity in the 0th moment map. 
However, we cannot definitively rule out larger values of the inclination (within the quoted error), and therefore lower values of v$_{\rm max}$.
The uncertainty quoted for v$_{\rm max}$ folds in the uncertainty in the inclination.  

\subsubsection{Dynamical Mass and the CO--to--H$_{2}$ conversion factor}
\label{Mdyn}

Assuming the parameters from the best fit model, we derive a dynamical mass for GN20 of 5.4 $\pm$ 2.4 $\times$ 10$^{11}$ M$_{\sun}$. 
The uncertainty was estimated assuming 1$\sigma$ uncertainties of 100 km s$^{-1}$ and 2 kpc on the rotational velocity and radius, respectively.
This estimate is based on dynamical modeling, making it more robust than previous estimates for this source \citep[which also relied on higher--J transition lines;][]{2009ApJ...694.1517D, 2010ApJ...714.1407C}.
We will now use this estimate to set limits on the CO--to--H$_{2}$ mass conversion factor.

In Section~\ref{H2mass} we calculated molecular gas masses assuming a CO--to--H$_{2}$ conversion factor of $\alpha_{\rm CO} = 0.8$ M$_{\sun}$ (K km s$^{-1}$ pc$^{2}$)$^{-1}$, the value derived for ULIRGs. 
The mass conversion factor is thought to vary with environment, however, depending on several factors including metallicity, excitation, and interstellar medium pressure \citep[][and references therein]{2008ApJ...686..948B}.
Generally, it is thought that it may decrease for objects with large gas surface densities \citep{1998ApJ...507..615D, 1997ApJ...484..702S, 2008ApJ...680..246T}, ranging from 0.8 M$_{\sun}$ (K km s$^{-1}$ pc$^{2}$)$^{-1}$ for ULIRGs up to $\sim$4.3 M$_{\sun}$ (K km s$^{-1}$ pc$^{2}$)$^{-1}$ for giant molecular clouds (GMCs) in the Milky Way.  
(All values stated here include Helium.)
While the ULIRG value of 0.8 M$_{\sun}$ (K km s$^{-1}$ pc$^{2}$)$^{-1}$ is often used for SMGs, there is not yet any firm evidence for what the SMG value should be. 
It is possible that for the more luminous SMGs, it is even lower. 
\citet{2008ApJ...680..246T} find that a Galactic conversion factor is strongly disfavored for their nine SMGs, with the lowest chi--squared values for their fits resulting from more ULIRG--like values. 
However, their resolution did only marginally resolve their sources, and they rely on CO(4--3) and even higher--order transitions that may be more centrally concentrated (see Section~\ref{excitation}).

Our derived dynamical mass allows us to put constraints on the conversion factor for GN20. 
If we assume that GN20 is composed of 100\% molecular gas, then we derive a conversion factor of $\alpha_{\rm CO}$ $=$ 3.3 $\pm$ 1.8 M$_{\sun}$ (K km s$^{-1}$ pc$^{2}$)$^{-1}$ to account for the total mass in the system. 
This is consistent (within the large uncertainties) with the Galactic value, but it represents the most extreme case.
In reality, the total mass will also include contributions from stars, dark matter, and dust.
The stellar component has been estimated to be 2.3 $\times$ 10$^{11}$ M$_{\sun}$ by \citet{2009ApJ...694.1517D} based on SED fitting to the ACS through IRAC photometry and assuming a \citet{2003ApJ...586L.133C} IMF. 
(Note that the estimate relies most heavily on the IRAC data, which is coincident with the CO emission.)
The contribution from dark matter is largely unknown, but observations of high--z galaxies suggest the value may be roughly 25\% \citep{2008ApJ...687...59G, 2010ApJ...713..686D}.
Using these estimates, and ignoring the contribution from dust, which is thought to make up only a tiny fraction of the total mass \citep[2 $\times$ 10$^9$ M$_{\sun}$;][]{2011ApJ...740L..15M}, 
we derive a conversion factor\footnote{Note that the formal uncertainty is larger but includes negative values of $\alpha_{\rm CO}$ and is therefore unphysical. Here, we have assumed an uncertainty of $\sim$50\%.}
 of $\alpha_{\rm CO}$ $=$ 1.1 $\pm$ 0.6 M$_{\sun}$ (K km s$^{-1}$ pc$^{2}$)$^{-1}$. 
This estimate is in agreement with the limit $\alpha_{\rm CO}$ $<$ 1.0 derived recently for GN20 by \citet{2011ApJ...740L..15M} using the local $M_{\rm gas}/M_{\rm dust}$ vs. metallicity relation.

\begin{figure*}[]
\centering
\includegraphics[scale=0.72]{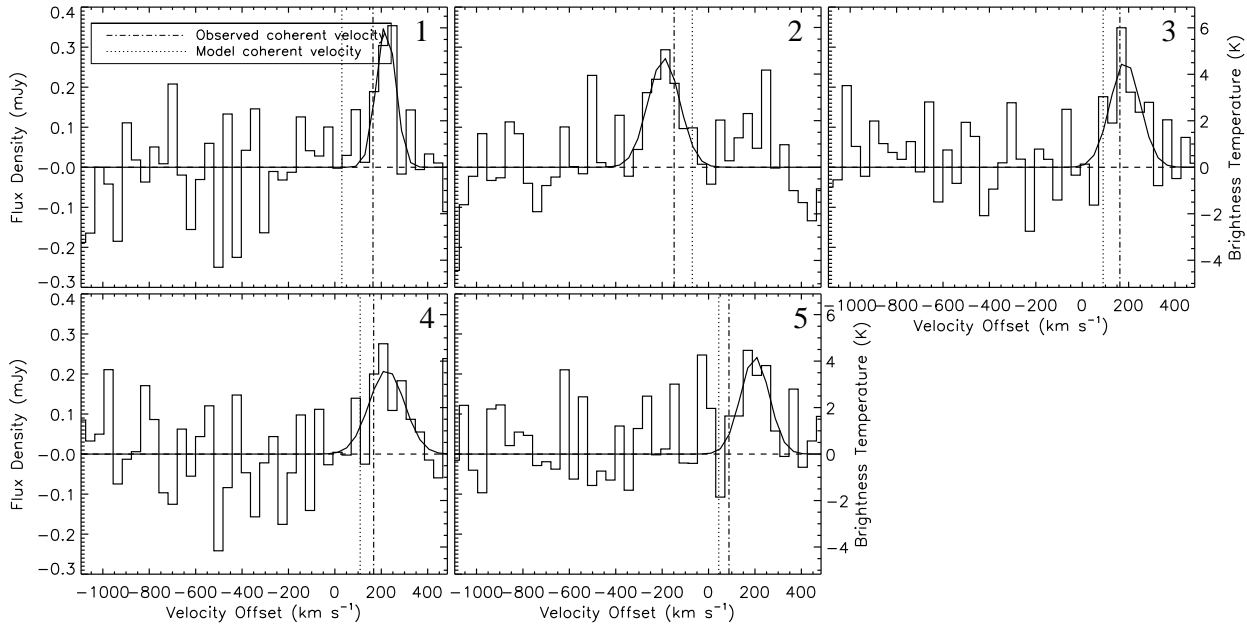}
\caption{CO(2--1) spectra of five molecular gas clumps in GN20. The spectra have been binned into channels of 39 km s$^{-1}$. The dot--dashed line shows the observed coherent velocity at the position of the clump, and the dotted line shows the model coherent velocity. The S/N of the spectra is not very high, but still allows us to derive approximate line widths for the individual clumps for the first time. The results for individual clumps are summarized in Table~\ref{tab-2}.}
\label{fig:clump_spectra}
\end{figure*}

\subsection{Molecular Gas Clumps}
\label{clumps}

\subsubsection{Clump search}

In order to study the properties of the molecular gas in GN20 in more detail, we have attempted to identify individual gas clumps in the disk of GN20.
We used two different methods to identify potential clumps,
judging as reliable only those clumps that were independently identified by both algorithms. 
The first method used the AIPS task SERCH.
SERCH is a ``matched-filter" analysis employing a Gaussian kernel and used in source--finding \citep[e.g.][]{2005MNRAS.362..609B, 2005A&A...429L..51K, Aravena_cluster}, and the algorithm is described in more detail in \citet{1991ApJ...377L..65U}.
We used a cube at our native angular resolution (0.19$^{\prime\prime}$) and with a spectral binning of 39 km s$^{-1}$.
We then searched for emission line sources in the three--dimensional data, optimizing the search for linewidths between 80--200 km s$^{-1}$ in steps of the channel size ($\sim$40 km s$^{-1}$).
A S/N cut of 4.5$\sigma$ yielded 11 positive peaks and two negative peaks in a generous 4$^{\prime\prime}$$\times$4$^{\prime\prime}$ region centered on GN20. 
Based on the number of negative detections, we can expect $\sim$2 of the positive peaks to be spurious. 
Ten of the positive peaks lie on the disk of GN20, as defined by the 0th moment map in Figure~\ref{fig:GN20_mom}, and one positive peak (and both negative peaks) lie off of the disk. 

To test the robustness of the clumps identified in GN20 by SERCH, we repeated the search using an independent method. The second method is built upon the principles of Bayesian inference, and we will briefly describe it here. (For a more detailed description, see Lentati et al. 2012, in preparation.) 
This method involves using the multimodal nested sampling algorithm MULTINEST \citep{2008MNRAS.384..449F} to efficiently sample the posterior distribution by fitting a simple parametric model to the data cube described 
both spatially and spectrally by a Gaussian. The spatial model parameters are fixed to match the clean beam, since the clumps are not expected to be resolved. Spectrally, the model is a Gaussian line profile described by a profile width $\sigma_{\rm \nu}$ and peak amplitude $A_{\rm p}$. The priors on $\sigma_{\rm \nu}$ and $A_{\rm p}$ are set as uniform between 20--150 km s$^{-1}$ and 0.0--0.4 mJy, respectively, and are considered independent. 

We then quantified the probability that the fitted model describes a real clump by using Bayesian model selection \citep[see e.g.][]{2003MNRAS.338..765H}.
The two models considered were:\\

\noindent (1) -- ``There is no clump present at that position"\\
(2) -- ``There exists a clump with $A_{\rm p}>0$ at that position"\\

We evaluated the evidence associated with the posterior for these competing models, where the evidence is the average of the likelihood over the prior. 
Specifically, MULTINEST returns the evidence and associated error for each peak in the posterior which can then be compared with the evidence for there being no clump present. 
The model corresponding to the posterior mode with the 
greatest probabiliy was then subtracted from the data, and the process was repeated until no sources above a low threshold probability were found.  
Individual clumps were identified by requiring a minimum distance of 0.12$^{\prime\prime}$ and two frequency channels between their centers.  
If multiple clumps were returned within that distance, the clump with the greatest evidence was chosen. 
This method returned a list of positive and negative clumps, sorted in order of their probability. 
The negative clumps were all very low on the list, with 20 positive clumps judged as more probable than the most probable negative clump. 

\begin{deluxetable*}{ l c c c c c c c }
\tabletypesize{\small}
\tablewidth{0pt}
\tablecaption{Molecular Gas Clumps Properties \label{tab-2}}
\tablehead{
\colhead{Clump\tablenotemark{a}} & \colhead{S$_{\rm CO(2-1)}$} & \colhead{FWHM$_{\rm CO(2-1)}$} & \colhead{S/N} & \colhead{T$_{\rm B}$ (rest frame)} & \colhead{$M(\rm H_2)$} & \colhead{\% Total $M(\rm H_2)$} & \colhead{$M_{\rm virial}$} \\
\colhead{} & \colhead{[$\mu$Jy]} & \colhead{[km s$^{-1}$]} & \colhead{[$\sigma$]} & \colhead{[K]} & \colhead{[$\times$ 10$^{9}$ ($\alpha_{\rm CO}/0.8)$ M$_{\odot}$]} & \colhead{[$\times$ 10$^{9}$ M$_{\odot}$]} & \colhead{[$\times$ 10$^{9}$ M$_{\odot}$]}}
\startdata
1	& 360 $\pm$ 100  & 90 $\pm$ 40 & 5.3 & 31 $\pm$ 8 & 4.4 $\pm$ 2.2 & 3\% $\pm$ 2\% & $<$1.0 $\pm$ 0.9\\
2	& 270 $\pm$ 60  & 150 $\pm$ 50 & 5.4 & 23 $\pm$ 5 & 6.0 $\pm$ 2.4 & 5\% $\pm$ 2\%  & $<$3.4 $\pm$ 2.2\\
3	& 270 $\pm$ 60  & 140 $\pm$ 40 & 5.1 & 23 $\pm$ 6 & 5.5 $\pm$ 2.1 & 4\% $\pm$ 2\% & $<$2.9 $\pm$ 1.8\\
4	& 180 $\pm$ 50  & 160 $\pm$ 60 & 4.5 & 16 $\pm$ 4 & 4.2 $\pm$ 1.9 & 3\% $\pm$ 1\% & $<$3.6 $\pm$ 2.6\\
5	& 240 $\pm$ 70 &  140 $\pm$ 50 & 4.6 & 21 $\pm$ 6 & 4.8 $\pm$ 2.3 & 4\% $\pm$ 2\% & $<$2.7 $\pm$ 2.0
\enddata
\tablenotetext{a}{See Figure~\ref{fig:GN20_clumps}.}
\end{deluxetable*}

We then cross--matched the lists of clump candidates produced by both SERCH and the Bayesian modeling,
requiring the clump centers returned by the two algorithms to agree within a beam radius in order to count as the same clump. 
Of the ten potential clumps identified on the disk of GN20 by SERCH, six were independently identified by the Bayesian modeling (and these were judged six of the seven most probable clumps by the Bayesian modeling).
The four SERCH clumps that were not identified by the Bayesian modeling included the two SERCH clumps furthest out on the disk, and two SERCH clumps which were closer than the minimum distance set in the Bayesian modeling.
Of the six common clump candidates, one was discarded 
because of an unnaturally thin profile (i.e.\ a single spectral channel).
The final sample therefore consists of five clumps. 
These are the most probable clumps which were independently identified by the two different methods, and therefore the five strongest clump candidates.
They are labeled in Figure~\ref{fig:GN20_clumps} at their positions on the CO(2--1) 0th (left) and 1st (right) moment maps with beam--sized circles (1.3 kpc).
As these clumps were identified in the three--dimensional data, they do not necessarily perfectly align with the most significant peaks in the integrated map.
We also note that, while these clumps were judged to be the most reliable clumps, they in no way consistute a comprehensive census of the clumps in GN20.

Spectra for these five clumps are shown in Figure~\ref{fig:clump_spectra}, where the panel numbers correspond to the clump numbers in Figure~\ref{fig:GN20_clumps}.  
Gaussian fits to the spectra are overplotted.
The coherent velocity at the position of each clump is shown in each panel, and is based both on the model velocity field at lower spatial resolution (dotted line) and the data velocity field at its native resolution (dot--dashed line).
In general, the clumps are within $\sim$100 km s$^{-1}$ of the (observed) coherent velocity at their position on the disk.
The fact that they are not centered exactly on the observed coherent velocity is due to the presence of other mass along the line--of--sight.
The model velocity field is based on data at a lower spatial resolution (0.77$^{\prime\prime}$) and so does worse (generally) at predicting the clump velocities.
While these spectra are low S/N, they allow us to estimate the properties of individual gas clumps in a z$=$4 galaxy for the first time.

\subsubsection{Clump Properties} 
\label{clump_props}

With this data, we attempt to estimate some basic properties of the molecular gas clumps.
We caution that this analysis is based on low S/N data and will need to be verified by even higher sensitivity observations.
To begin with, we use the spectral information to determine brightness temperatures of individual clumps without diluting the values by averaging in velocity.
Derived brightness temperatures (using the Rayleigh--Jeans approximation) range from 3.2--6.2 K and can be read off of the right--hand y--axis in Figure~\ref{fig:clump_spectra}.  
After correcting to the rest--frame, values for individual clumps range from $\sim$16 K up to 31 K (Table~\ref{tab-2}).
For reference, Planck temperatures (i.e.\ the temperatures calculated using the full blackbody instead of just the Rayleigh--Jeans approximation) are typically $\sim$20\% higher.

Since the clumps appear to be unresolved on the scales probed by our observations (1.3 kpc), the brightness temperature values are most likely lower limits.  
Nevertheless, it is interesting to note that the peak brightness temperatures derived for the clumps are already approaching (or even exceeding) the dust temperature \citep[33K;][]{2011ApJ...740L..15M}.
If we instead compare to the large velocity gradient \citep[LVG; e.g.][]{1974ApJ...187L..67S} gas model fit by \citet{2010ApJ...714.1407C}, where the best--fit model consisted of a lower and a higher excitation component, the higher excitation component still has a kinetic temperature of only 45 K.
The change in surface area required to make up this temperature difference is small.
This may indicate that we are close to resolving the clumps, though we cannot make this statement any more quantitative with the current dataset.


We next fit the linewidths of the clumps.
There are several factors that contribute to broadening the clump linewidths beyond their intrinsic values. 
One factor is the instrumental velocity resolution, which we have removed in quadrature from the measured linewidths. 
Another factor is the rotational contribution from the large--scale velocity structure of the galaxy.
To estimate this contribution, we have taken the model velocity field and measured the velocity gradient across a beam--sized area centered on each clump, dividing by two to approximate the weighting of the beam.
The resulting linewidths are listed in Table~\ref{tab-2} and have a median value of 140 km s$^{-1}$.
The errors were derived from the Gaussian fits and assuming a 30\% error on the measured velocity gradient.

Using the Gaussian fits to the spectra in Figure~\ref{fig:clump_spectra}, we determined the molecular gas masses of the clumps (Table~\ref{tab-2}).
For this, we used the same relations and assumptions as in Section~\ref{H2mass}, including an H$_2$--to--CO conversion factor of $\alpha_{\rm CO}$$=$0.8 M$_{\sun}$ (K km s$^{-1}$ pc$^{2}$)$^{-1}$.
We estimate that the clumps have H$_2$ masses of $\sim$5 $\times$ 10$^{9}$ $\times$ $(\alpha_{\rm CO}/0.8)$ M$_{\sun}$, or a few percent of the total gas mass each.

\begin{figure*}[]
\centering
\includegraphics[scale=0.63]{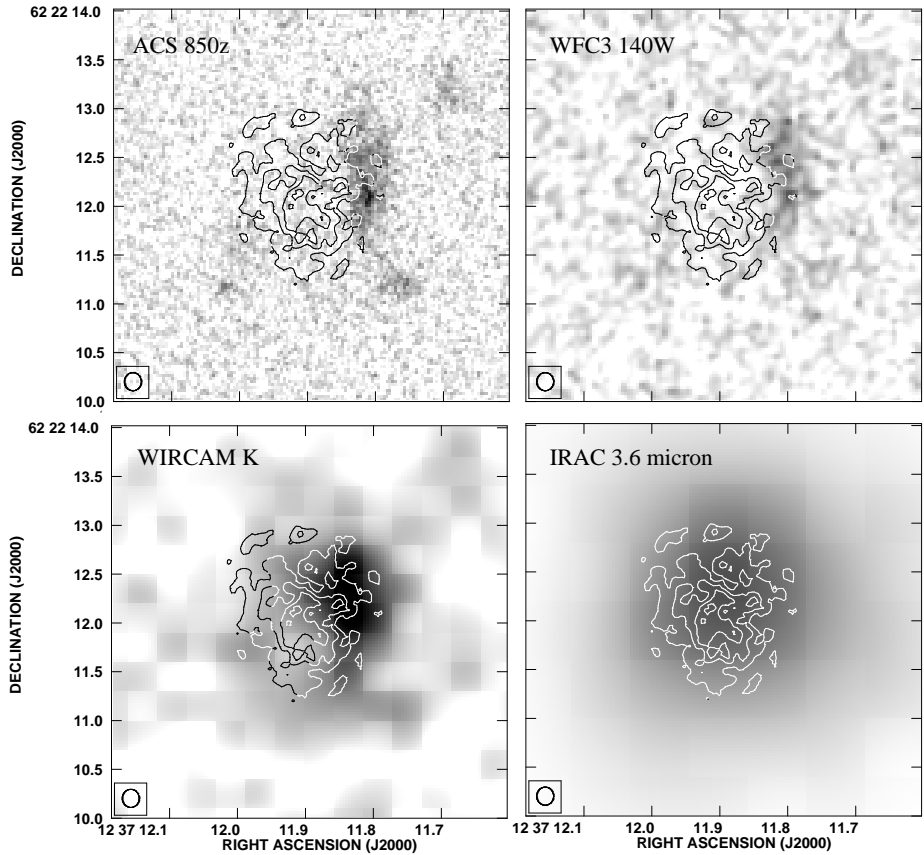}
\caption{CO(2--1) contours at intermediate resolution overlaid on a selection of multiwavelength data. The labels in the top left corners refer to the greyscale images -- see text for details.}
\label{fig:GN20_multiwav}
\end{figure*}

Taking the beam size as an upper limit on the clump size, the mass surface densities of the clumps are $>$3,200--4,500 $\times$ ($\alpha_{\rm CO}/0.8$) M$_{\odot}$ pc$^{-2}$.
Similar surface densities have been seen in some compact SMGs with double--peaked line profiles \citep[3,000--10,000 M$_{\odot}$ pc$^{-2}$;][]{2010ApJ...724..233E}.
GMCs in the nearby universe, on the other hand, are typically several orders of magnitude (in volume) smaller and have surface densities approximately 20 times lower \citep{1987ApJ...319..730S}.
\citet{2008ApJ...680..246T} argue that a major merger is necessary to produce such high mass surface densities as those seen in SMGs. 
However, \citet{2010ApJ...724..233E} have pointed out that this argument alone is insufficient to distinguish a merger.
Indeed, if we assume that the clumps are spherical and roughly the size of the beam (at maximum), then we derive (average) volume densities of only 70--110 cm$^{-3}$,
consistent with the mean density of local GMCs \citep{1987ApJ...319..730S}.  
The dense cores of GMCs show much higher values, $\sim$10$^{4}$ cm$^{-3}$ \citep[e.g.][]{1997ApJ...491..615G}, and the same may also be true for the cores of the clumps in GN20.
The mean H$_2$ volume density of GN20 as a whole is only $\sim$15 cm$^{-3}$.

We estimated the virial masses of the clumps using the isotropic virial estimator \citep[e.g.][]{2006ApJ...646..107E}:
\begin{equation}
M_{\rm virial} = \frac{C\sigma_{\rm v}^{2}R_{\rm g}}{G} (M_{\odot})
\label{eqn1}
\end{equation}
where C $=$ 5 is the dimensionless prefactor for a uniform sphere, 
$\sigma_{\rm v}$ is the observed 1D velocity dispersion in km s$^{-1}$, $R_{\rm g}$ is the graviational radius, and $G = 1/232$ pc (km/s)$^{2}$ M$_{\odot}^{-1}$ is the gravitational constant. 
If we take the beam HWHM as an upper limit on the gravitational radius,
we estimate virial masses for the clumps of $<$1.0--3.6 $\times$ 10$^{9}$ M$_{\odot}$ (Table~\ref{tab-2}). 
These masses are all within 1--2$\sigma$ of the observed molecular gas masses, even assuming a low ULIRG-like mass conversion factor of $\alpha_{\rm CO}$ $=$ 0.8 M$_{\sun}$ (K km s$^{-1}$ pc$^{2}$)$^{-1}$ and ignoring the possible presence of significant stellar mass.
Within the considerable uncertainties, it is possible that the clumps have masses consistent with self--gravitation.

The gas masses derived for the clumps are already on the high side (in general) of the virial mass estimates,
even with a low conversion factor and despite the fact that the virial mass estimates are upper limits.
This implies that for the majority of the clumps, the gas masses would significantly overshoot the virial masses for $\alpha_{\rm CO}$ $>>$ 0.8 M$_{\odot}$ (K km s$^{-1}$ pc$^{2}$)$^{-1}$.
Assuming the clumps are entirely composed of molecular gas and using $M_{\rm virial}/L'_{\rm CO(2-1)}$ as an estimate of $\alpha_{\rm CO}$, we find $\alpha_{\rm CO}$ for the clumps is $<$0.2--0.7 M$_{\odot}$ (K km s$^{-1}$ pc$^{2}$)$^{-1}$. 


Recently, a couple of other observational studies have looked at molecular gas clumps in $z>1$ star forming galaxies.
For example, \citet{2010Natur.463..781T} observed molecular gas clumps in a `normal' star forming galaxy at $z \sim 1.2$.
These clumps had gas masses of $\sim$5$\times$10$^{9}$ M$_{\odot}$, diameters $<$2--4 kpc, brightness temperatures of $\sim$10--25 K, and gas volume densities $\sim$100 cm$^{-3}$, 
similar to what we get for GN20 if we assume that the clumps are roughly the beamsize.
Their velocity dispersions, however, were only $\sim$20 km s$^{-1}$,
causing them to 
postulate that the `clumps' observed were more likely loose conglomerates of several GMCs.
Another recent paper by \citet{2011ApJ...742...11S} identified molecular gas clumps in a lensed star forming galaxy at $z=2.3$, SMM J2135. 
The clumps in SMM J2135 had gas masses of $\sim$0.3--1$\times$10$^{9}$ M$_{\odot}$, a median diameter of 200 pc, and gas volume densities of $\sim$1,000--40,000 cm$^{-3}$.
To achieve volume densities this high, the clumps in GN20 would need to be up to an order of magnitude smaller than the current beamsize.

\begin{figure*}[]
\centering
\includegraphics[scale=0.7]{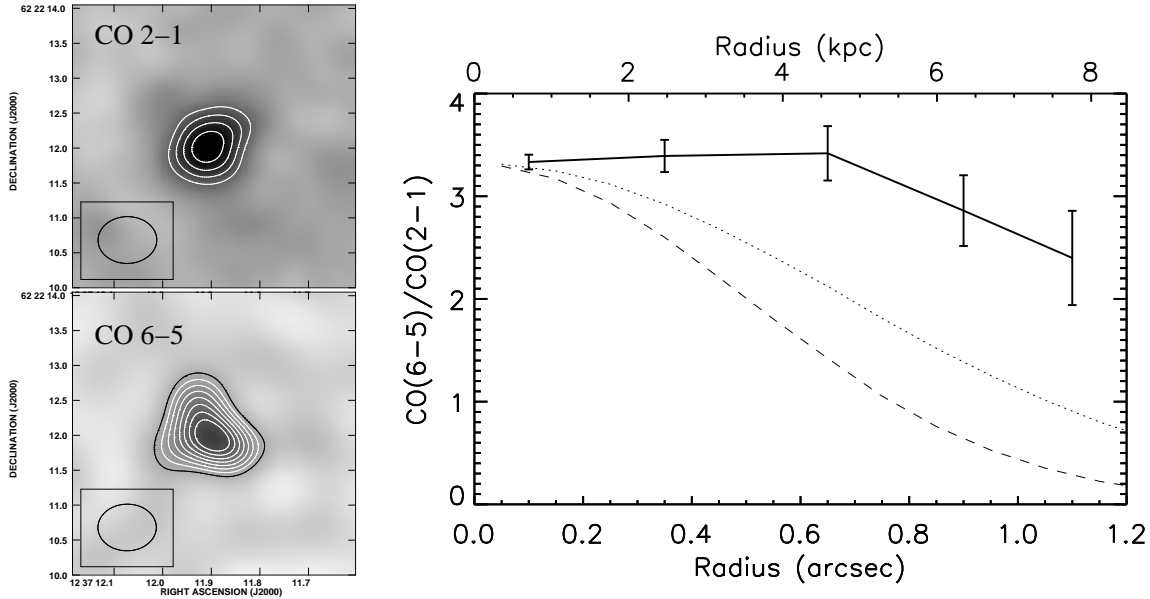}
\caption{Spatially--resolved molecular gas excitation in GN20. The left panels show the CO(2--1) and CO(6--5) images at the same resolution (0.84$^{\prime\prime}$ $\times$ 0.67$^{\prime\prime}$) and beam position angle. The right plot shows the result of dividing the CO(6--5) image by the CO(2--1) image (error bars do not include overall flux calibration uncertainties). Noise dominates the division at radii larger than $\sim$1.2$^{\prime\prime}$. The dotted and dashed lines are models (at the same resolution) where the CO(6--5) emission profile has an exponential scale length that is 2 and 5 times more compact than the CO(2--1), respectively. These model images have been normalized by the peak value in the image and are meant to demonstrate that changes in the excitation ratio are detectable at this resolution.}
\label{fig:GN20_excitation}
\end{figure*}

\subsection{Multiwavelength Comparison}
\label{multiwav}

Figure~\ref{fig:GN20_multiwav} shows the 0th moment CO(2--1) contours at intermediate resolution overlaid on a selection of the multiwavelength data available for GN20, including the HST$+$ACS z850--band \citep[top left;][]{2004ApJ...600L..93G}, 
the WFC3 140W--band (top right), 
the WIRCAM K--band (bottom left), 
and the IRAC 3.6$\mu$m data (bottom right).
Not shown is the higher--resolution VLA$+$MERLIN data \citep{2009MNRAS.400..670C}, which was compared to GN20 in CO in \citet{2010ApJ...714.1407C}. 
The object detected in the HST$+$ACS z850--band data (top left panel) was followed up spectroscopically with the Deep Imaging Multi--Object Spectrograph (DEIMOS) on Keck II and determined to have a redshift of z$=$4.055 \citep{2009ApJ...694.1517D}, consistent with the redshift derived for GN20 from its CO lines.

Assuming the object is related to GN20, as indicated by the good match in redshift, then the question is what is causing the clear offset from the CO and radio/(sub)millimeter position also reported in previous work on GN20 \citep[for e.g.,][]{2010ApJ...714.1407C, 2006MNRAS.370.1185P, 2008ApJ...688...59Y, 2006ApJ...640L...1I}. 
We have confirmed that the astrometry of the HST imaging is good to $<$0.15$^{\prime\prime}$ \citep{2007ApJ...670..156D}, so astrometric error is unlikely. 
In addition, the counterpart in the WFC3 140W--band imaging is similar in position/morphology. 
Such large offsets between dust/molecular gas emission and (rest--frame) UV/optical imaging have been seen in a number of SMGs \citep{2005ApJ...622..772C, 2008ApJ...681L..53C, 2010ApJ...720L.131R}. 
A possible hint in this case comes from the WIRCAM K--band image, which is similarly offset, but shows extended emission in the direction of the radio/(sub)millimeter counterparts. 
This extended emission is significant at the $\sim$6$\sigma$ level.
A progression through the observations shows that the peak of emission shifts toward the radio position as we sample longer wavelengths. 
One interpretation is therefore that the radio/(sub)millimeter emission is offset from the (observed) optical emission because the dust that is presumably associated with the molecular gas heavily obscures the rest--frame UV and optical emission. 
A significant amount of dust obscuration is not surprising in a starburst of several thousand M$_{\sun}$ yr$^{-1}$, the SFR estimated for GN20 \citep{2011ApJ...740L..15M, 2009ApJ...694.1517D}. 
Heavy dust obscuration is thought to explain the complete absence of optical counterparts for some SMGs \citep[e.g.,][]{2012Natur.486..233W}.
The top left panel in Figure~\ref{fig:GN20_multiwav} shows that the rest--frame UV emission is well aligned with the edge of the 0th moment CO contours, giving the clearest indication that the ACS counterpart \textit{is} physically related but is simply offset due to differential dust obscuration.

\subsection{Spatially--resolved Gas Excitation}
\label{excitation}

To investigate the spatially--resolved molecular gas excitation, we compare our CO(2--1) data with the CO(6--5) Plateau de Bure Interferometer (PdBI) data from \citet{2010ApJ...714.1407C}.  The CO(6--5) data are uniformly--weighted at an angular resolution of 0.84$^{\prime\prime}$ $\times$ 0.67$^{\prime\prime}$.  We tapered the CO(2--1) data to the same resolution and imaged over the same velocity range (800 km s$^{-1}$).  

We then used the AIPS task IRING to calculate the average CO(6--5)/CO(2--1) ratio as a function of radius.  The centroids of the CO(6--5) and CO(2--1) emission peaks were consistent within 0.08$^{\prime\prime}$, so we used the average of the positions as the center of the concentric annuli.  
We set the width of the annuli to be 0.25$^{\prime\prime}$, i.e., one--third of the synthesized beam, and we specified a major axis position angle of 115$^{\circ}$ and an inclination of 30$^{\circ}$ as determined in Section~\ref{dyn}.

The results are shown in Figure~\ref{fig:GN20_excitation}.  
The left panels show the CO(2--1) and CO(6--5) emission, using the native beamshape of the CO(6--5) map.
The plot on the right is the result of dividing the CO(6--5) emission by the CO(2--1) emission, averaging in elliptical annuli as explained above. 
Note that the error bars are purely statistical and do not account for the overall flux calibration uncertainties in the maps (which in both cases are of order 10\%).

The CO(6--5)/CO(2--1) ratio is an indicator of the excitation state of the gas. 
The ratio hovers around a value of $\sim$3,
a factor of three lower than expected for thermal excitation.
This indicates that the CO(6--5) line is sub--thermally excited, consistent with the CO excitation ladder shown in \citet{2010ApJ...714.1407C}.
Carilli et al.\ determined that the data were best fit with a two--component gas model including a diffuse, low--excitation component and a more concentrated high--excitation component. 

With the data at hand, we can go one step further and investigate the radial dependence of the CO ratio. 
From Figure~\ref{fig:GN20_excitation}, we see that the ratio is consistent with a constant value out to at least 1$^{\prime\prime}$, corresponding to the observed radius of the CO(2--1) emission in the 0th moment map. 
Note that we have spaced the radial bins so as to achieve Nyquist sampling, so the first two data points are within the beam.
Nevertheless, this result tells us that to first order, the CO(6--5) emission has the same radial profile as the CO(2--1) emission. 
This would not be the case if, for example, the CO(6--5) emission (which is more closely related to star formation) was more tightly concentrated than the lower--excitation CO(2--1) emission.
To illustrate this point, we show two different models (at the same resolution as the data) where the CO(6--5) emission is more centrally--concentrated than the CO(2--1) emission. 
In the first model (dotted line), the exponential scale length of the CO(6--5) emission is two times smaller than that of the CO(2--1) emission.
In the second model (dashed line), the CO(6--5) emission has a scale length that is five times smaller.
In both cases, the CO(2--1) disk parameters come from our dynamic modeling in Section~\ref{dyn}.
The models have been normalized to the peak value in the maps and are meant to demonstrate the clear drop--off in the excitation ratio that would be detectable for a central star forming event even at this resolution.
Instead, the flatness we observe is consistent with the idea that the two transitions are cospatial, and that the star formation is therefore spread out over a large portion of the (14 kpc diameter) disk.

\section{WHAT IS DRIVING THE STARBURST?}
\label{discussion}

GN20 has several properties which make it unique as an SMG. 
It has a higher redshift than the typical SMG,
and it is extremely bright -- 
its 850 $\mu$m flux density of 20.3 mJy makes it 
one of the most luminous starburst galaxies known 
\citep{2006MNRAS.370.1185P}.
In addition, it appears to be part of a massive proto--cluster of galaxies:
there are two other $z\sim4.05$ SMGs just $\sim$25$^{\prime\prime}$/170 kpc away known as GN20.2a and GN20.2b \citep{2009ApJ...694.1517D, 2011ApJ...739L..33C};
there are 14 $z\sim4$ Lyman Break Galaxies within 25$^{\prime\prime}$;
and there is an overdensity of $z > 3.5$ IRAC selected galaxies in the field.

The question is: 
what process is driving the gas accretion and conversion into stars in this example of one of the earliest extreme starburst galaxies?
The majority of SMGs are thought to be gas--rich major mergers.
There are multiple examples of SMGs with very disturbed kinematics and 
even showing multiple components,
providing direct observational evidence for the merger picture \citep[e.g.,][]{2008ApJ...680..246T, 2010ApJ...724..233E, 2011MNRAS.412.1913I,2011ApJ...733L..11R, 2011ApJ...739L..31R}.
For example, by identifying likely radio and/or mid--IR identifications to SMGs in the SHADES Source Catalog \citep{2006MNRAS.372.1621C}, 
\citet{2007MNRAS.380..199I} found that significantly more SMGs have multiple robust counterparts than would be expected by chance alone.

In GN20, on the other hand, we observe an extended, rotating gas disk with a velocity field that is consistent with a flat rotation curve.
The gas reservoir appears to be fragmented into multiple clumps, with each clump making up only a few percent of the total mass.
This morphology stands in contrast to prototypical merging sources like Arp 220 and other ULIRGs, where the molecular gas is typically concentrated in one or two dense, centrally--located regions.  
If the diffuse and clumpy gas components correspond to the two gas excitation states predicted to exist in GN20 with LVG modeling \citep{2010ApJ...714.1407C}, then the clumpy component (presumably given by the higher excitation state gas) only makes up $\sim$half of the total gas mass, with the remainder existing as extended, diffuse emission.



Recently, cold mode accretion (CMA) has been suggested as a possible alternative to the major merger formation scenario for some SMGs \citep{2001astro.ph..7290F, 2006ApJ...639..672F, 2010MNRAS.404.1355D,2010ApJ...714.1407C}. 
In this scenario, the star formation in these SMGs would be instead fueled by minor mergers and the smooth infall of cold gas along streams.
Unlike a major merger, this process can leave the disk relatively undisturbed, especially for a galaxy as massive as GN20 \citep{2010MNRAS.404.1355D}. 
However, wet mergers have been shown to rapidly relax into smooth disks, so the presence of ordered rotation alone cannot immediately be taken as an argument against a major merger \citep{2006ApJ...645..986R, 2008ApJ...685L..27R, 2009ApJ...694L.158B}. 

If GN20 were a major merger where the gas had re--virialized into a rotating disk, one may ask whether it would still be an SMG at this late stage?  
According to 
hydrodynamic simulations of major mergers,
the submm--bright phase in a ($z\sim2$) major merger--driven starburst should occur as the galaxies approach final coalescence (\citealt{2009MNRAS.400.1919N}, but c.f. \citealt{2011ApJ...743..159H, 2012arXiv1203.1318H}).
During this brief phase ($\sim$0.03 Gyr), the 850 $\mu$m flux is significantly boosted and the galaxy may be selected as a luminous submillimeter source. 
However, 
according to the simulations, 
this phase coincides with the stage when the disk--like morphology is most disturbed, with the largest fraction of gas tidally disrupted from disk rotation 
\citep[Figure 6]{2009MNRAS.400.1919N}.  
These simulations also show that the gas during the SMG phase is relatively concentrated, with a characteristic radius of $\sim$1.5 kpc.
Such a compact morphology is expected, since the large--scale gravitational torques induced by gas--rich major mergers
are efficient at removing angular momentum \citep{1996ApJ...471..115B}.
In GN20, on the other hand, we simultaneously observe 1) an extremely bright (S$_{850}$ $=$ 20.3 mJy) submillimeter flux, 
2) ordered rotation, and
3) a very extended (14 kpc) gas disk.
Taken together, these observations may be an indication that 
the star formation in GN20 is
fueled by some process other than a major merger.
A similar argument was put forward by \citet{2010MNRAS.405..219B} regarding the SMG HDF132.
Of course, this argument rests on the assumption that these simulations
provide an accurate representation of merger--driven SMG formation for all SMGs,
which may not be the case,
especially for notable SMGs like GN20.

On the other hand, there are certain properties of GN20 which may be difficult to explain without a major merger.
In particular, the rotational velocity derived from the dynamical modeling is extremely high (576 km s$^{-1}$). 
While we cautioned that this is largely dependent on the galaxy's inclination, (which is hard to constrain at our S/N),
a rotational velocity this large may be hard to reproduce in a spinning disk embedded in its dark matter halo, and the linewidth could be more easily explained as resulting from the final coalescence stage of a major merger.
Also noteworthy is the observation that GN20 is undergoing an extremely dusty massive starburst. 
This is supported by its high specific SFR \citep{2009ApJ...694.1517D}, and also by its hot FIR SED \citep{2011ApJ...740L..15M} and low CO--to--H$_2$ conversion factor, which both imply a high star formation efficiency.
The critical question is: is it possible to form such an object without relying on a major merger?
Locally, at least, the answer is no -- such objects are all major mergers (ULIRGs).

Perhaps the real answer lies somewhere in between the two scenarios.
\citet{2006ApJ...639..672F} studied $z\sim4$ B-band dropout galaxies in cosmological hydrodynamic simulations and found two instances of galaxies forming stars at over 1000 M$_{\odot}$ yr$^{-1}$, several times that of their 100 Myr averages.
These objects both resided in highly overdense regions resembling the cores of massive clusters.
The authors found that, while a merger of two massive galaxies was not the dominant mechanism for fueling star formation, there were still signs of interaction with many of the smaller nearby galaxies.
GN20 lies in a similarly dense environment -- the overdensity of B-band dropouts and IRAC--selected galaxies suggests a protocluster of total mass 10$^{14}$ M$_{\odot}$ centered on GN20 and the nearby SMGs GN20.2a and GN20.2b \citep{2009ApJ...694.1517D}.
It is likely, therefore, that even if there is a dominant rotating disk, enhanced star formation over the disk is being triggered by the gravitational torques from the other galaxies.

\section{CONCLUSIONS}
\label{conclusions}

We have presented VLA observations of CO(2--1) in the $z=4.05$ SMG GN20. 
These high--resolution data allow us to do detailed imaging of the molecular gas on scales down to 1.3 kpc. 
We use these data to model the overall gas dynamics and determine the properties of individual star forming clumps in an extreme starburst galaxy just 1.6 Gyr after the Big Bang.
We summarize our main findings in the following.

The data reveal a clumpy, extended gas reservoir in unprecedented detail. 
A carefully constructed 0th moment map recovers emission over an area 14 $\pm$ 4 kpc in diameter, and the first moment map shows a clear velocity gradient.
A dynamical analysis shows that the data are consistent with a large, rotating disk with a maximum rotational velocity of $v_{\rm max}$ $=$ 575 $\pm$ 100 km s$^{-1}$, an inclincation of 30$^{\circ}$ $\pm$ 15$^{\circ}$, and a dispersion of 100 $\pm$ 30 km s$^{-1}$.
While it is difficult to determine the exact shape of the rotation curve, 
a flat rotation curve gives the best fit to the data.

From the dynamical modeling, we derive a dynamical mass for GN20 of 5.4 $\pm$ 2.4 $\times$ 10$^{11}$ M$_{\odot}$.
We use this dynamical mass in combination with the stellar mass to put constraints on the CO--to--H$_2$ mass conversion factor ($\alpha_{\rm CO}$), 
finding $\alpha_{\rm CO}$ $=$ 1.1 $\pm0.6$ M$_{\sun}$ (K km s$^{-1}$ pc$^{2}$)$^{-1}$.
This value is consistent with a low, ULIRG--like value, 
although better data is needed to confirm this result.
Our estimate is also in agreement with recent limits set for GN20 using the local $M_{\rm gas}/M_{\rm dust}$ vs. metallicity relation.

We use two different source--finding techniques to identify five distinct molecular gas clump candidates on the disk of GN20.
These clumps have masses of a few percent of the total gas mass and a median linewidth of 140 km s$^{-1}$ (FWHM).
Their surface densities are $>$3,200--4,500 $\times$ ($\alpha_{\rm CO}/0.8$) M$_{\odot}$ pc$^{-2}$, corresponding to volume densities of $>$100 cm$^{-3}$ (where the lower limit corresponds to GMCs in the Milky Way).
Their peak brightness temperatures (16--31 K) are approaching both the average dust temperature (33K) and the `warm' component of a two--component LVG model, implying 
that we may be close to resolving the clumps.
Within the substantial uncertainties, all clumps are consistent with being self--gravitating masses,
and we constrain their CO--to--H$_2$ mass conversion factor to be $<$0.2--0.7 M$_{\sun}$ (K km s$^{-1}$ pc$^{2}$)$^{-1}$.


We compare the observed distribution of CO to the emission seen at other wavelengths.
This multiwavelength comparison demonstrates that the molecular gas is concentrated in a region of the galaxy which is heavily obscured in the rest--frame UV and optical. The optical counterpart likely constitutes a small percentage of the stellar light.

Using previous observations of the CO(6--5) in GN20, we examine the CO(6--5)/CO(2--1) excitation ratio in elliptically--averaged annuli.
The excitation ratio is consistent with a constant value over the extent of the 14 kpc disk.
This result implies that the star formation is spread out over a large portion of the disk, rather than concentrated in a central star forming event.

We discuss our results in the context of different fueling scenarios for GN20.
While the presence of an extended, rotating disk during the SMG phase may point toward a process other than a major merger (e.g., cold mode accretion), its large linewidth and extremely dusty starburst would be more easily explained in a merger picture.
If the star formation is fueled primarily by CMA, then, due to GN20's location in a dense protocluster environment, it is likely that it is being enhanced by interactions with nearby galaxies.


We conclude by highlighting the importance of low--J CO studies, which trace the bulk of the molecular gas, in studying the formation of massive galaxies throughout cosmic time.
Such studies cannot currently be done in a statistical way due to the time required to attain sufficient S/N at high--resolution, even with state--of--the--art instruments such as the VLA.
This fact makes targeted case studies such as this that much more important.


\acknowledgements
The authors wish to thank the anonymous referee for helpful comments which improved this paper.
We thank Glenn Morrison, Mark Dickinson, Bram Venemans, Gabriel Brammer, Desika Narayanan, Benjamin Weiner, and Dario Colombo for useful comments and discussions.
CC thanks the Kavli institute of Cosmology for their hospitality.
DR acknowledges funding from NASA through a Spitzer Space Telescope grant. 
The National Radio Astronomy Observatory is a facility of the National Science Foundation under cooperative agreement by Associated Universities, Inc.
This research has made use of the GIPSY package.


\bibliographystyle{apj}
\bibliography{Hodge}

\end{document}